\documentclass[conference]{IEEEtran}
\usepackage{cite}
\usepackage{graphicx}
\usepackage{amsmath,amssymb,amsfonts}
\usepackage{algorithmic}
\usepackage{textcomp}
\usepackage{multirow}
\usepackage{booktabs}
\usepackage{adjustbox}
\usepackage{array}
\usepackage{xcolor}
\usepackage{lscape}

\begin{document}

\title{When IoT Meet LLMs: Applications and Challenges }

\author{
\IEEEauthorblockN{İbrahim KÖK}
\IEEEauthorblockA{Department of AI and Data Engineering\\
Ankara University\\
Ankara, Turkey\\
Email: ikok@ankara.edu.tr}
\and
\IEEEauthorblockN{Orhan DEMİRCİ}
\IEEEauthorblockA{Department of Computer Engineering\\
Hacettepe University\\
Ankara, Turkey\\
Email: orhandemirci@hacettepe.edu.tr}
\and
\IEEEauthorblockN{Suat ÖZDEMİR}
\IEEEauthorblockA{Department of Computer Engineering\\
Hacettepe University\\
Ankara, Turkey\\
Email: ozdemir@cs.hacettepe.edu.tr}
}

\maketitle

\begin{abstract}
Recent advances in Large Language Models (LLMs) have positively and efficiently transformed workflows in many domains. One such domain with significant potential for LLM integration is the Internet of Things (IoT), where this integration brings new opportunities for improved decision making and system interaction. In this paper, we explore the various roles of LLMs in IoT, with a focus on their reasoning capabilities. We show how LLM-IoT integration can facilitate advanced decision making and contextual understanding in a variety of IoT scenarios. Furthermore, we explore the integration of LLMs with edge, fog, and cloud computing paradigms, and show how this synergy can optimize resource utilization, enhance real-time processing, and provide scalable solutions for complex IoT applications. To the best of our knowledge, this is the first comprehensive study covering IoT-LLM integration between edge, fog, and cloud systems. Additionally, we propose a novel system model for industrial IoT applications that leverages LLM-based collective intelligence to enable predictive maintenance and condition monitoring. Finally, we highlight key challenges and open issues that provide insights for future research in the field of LLM-IoT integration.

\end{abstract}

\section{Introduction}
In recent years, advancements in Artificial Intelligence (AI) have paved the way for smarter, more adaptable, and multimodal systems across various domains. Specifically, Generative AI and Large Language Models (LLMs) have emerged as groundbreaking and trending areas in AI, revolutionizing the field with their capabilities in complex reasoning, deep understanding, and generating diverse types of data \cite{kenton2019bert}. However, the success of these systems heavily relies on training advanced models using large and heterogeneous data sources. Beyond the game-changing power brought by current AI systems, significant advancements are needed in areas such as cost, trust, explainability, accountability, and ethics. Specifically, in the context of LLMs, challenges include data biases, lack of up-to-date information, privacy concerns, high computational power and energy consumption, memory and storage requirements, generation of false or fabricated information, limitations in complex reasoning, the need for task specificity, and risks associated with producing misleading content \cite{alkaissi2023artificial}. Addressing these issues and overcoming the current challenges, particularly in solving data-driven problems, holds great promise for the future. 

Internet of Things (IoT), a newly emerging technology, serves as a key data provider for the field of AI. The integration of LLMs and IoT systems create a synergistic relationship, offering mutual benefits for both technologies. On one hand, IoT provides LLMs with access to real-world, domain-specific data through its vast network of sensors and devices \cite{an2024iot,xu2024penetrative}. This data helps mitigate a critical challenge in LLMs—hallucination, where models generate contextually or physically implausible outputs \cite{huang2023survey}. By acting as their "sensing body parts," IoT devices feed real-time sensory inputs such as temperature, motion, and spatial data into LLMs, enabling them to enhance their understanding and produce more accurate, grounded outputs. This integration transforms LLMs into more perceptive and context-aware systems.

On the other hand, IoT systems stand to benefit significantly from the reasoning and decision-making capabilities of LLMs. IoT faces challenges such as poor human-device interactions, managing heterogeneous and large-scale data streams, and making real-time, complex decisions under resource constraints \cite{mekuria2021smart}. LLMs provide advanced reasoning frameworks to address these issues, enabling IoT systems to analyze data more effectively, resolve ambiguities in heterogeneous environments, and adapt dynamically to changing conditions \cite{hassanin2025pllm,cui2024llmind}.
Despite these benefits, there are challenges to LLM-IoT integration, primarily stemming from the large size of LLM models, which impose significant hardware constraints and bandwidth requirements \cite{friha2024llm}. These limitations necessitate innovative deployment strategies, such as resource management optimizations, distributed LLM into devices and servers, model optimization techniques, to ensure efficient integration without compromising performance \cite{gao2024dlora,yang2024perllm,wu2024fedbiot}.

In this paper, we investigate the integration of LLMs with IoT, focusing on how this synergy can enhance decision-making, optimize network performance, and offer scalable solutions for complex IoT systems. We also examine the challenges posed by data processing, real-time responses, and the security of IoT data. Through this exploration, the study contributes to the ongoing development of Generative IoT, offering new insights into its future impact on AI and IoT fields. The main contributions of this paper are as follows:
\begin{itemize}
    \item We examine the potential of integrating LLMs with IoT, investigating the development of the Generative IoT concept and the opportunities provided by this integration.
    \item We demonstrate how LLMs contribute to optimizing IoT networks and improving coordination between devices by breaking down tasks into smaller components.
    \item We summarize existing literature to provide insights into future directions, offering perspectives on how the integration of LLMs and IoT can evolve and address emerging challenges.
\end{itemize}

The rest of the paper is organized as follows: Section II present a brief background on LLMs. Section III explains LLMs and IoT integration.   Section IV introduces LLM-enabled IoT network architectures. Section V explores LLM-supported IoT applications in the literature. Section VI introduces the proposed LLM supported IoT system model. Section VII discusses future directions. Finally, Section VIII concludes the paper.

\section{Brief Background on LLMs}

LLMs are advanced language models based on the Transformer architecture, which uses a self-attention mechanism to capture long-range dependencies and understand word context efficiently. Unlike RNNs or LSTMs, Transformers handle large datasets with parallel processing, making them suitable for complex NLP tasks requiring deep contextual understanding and scalability \cite{vaswaniattention}.
\begin{figure*}
    \centering
   \includegraphics[scale=.56]{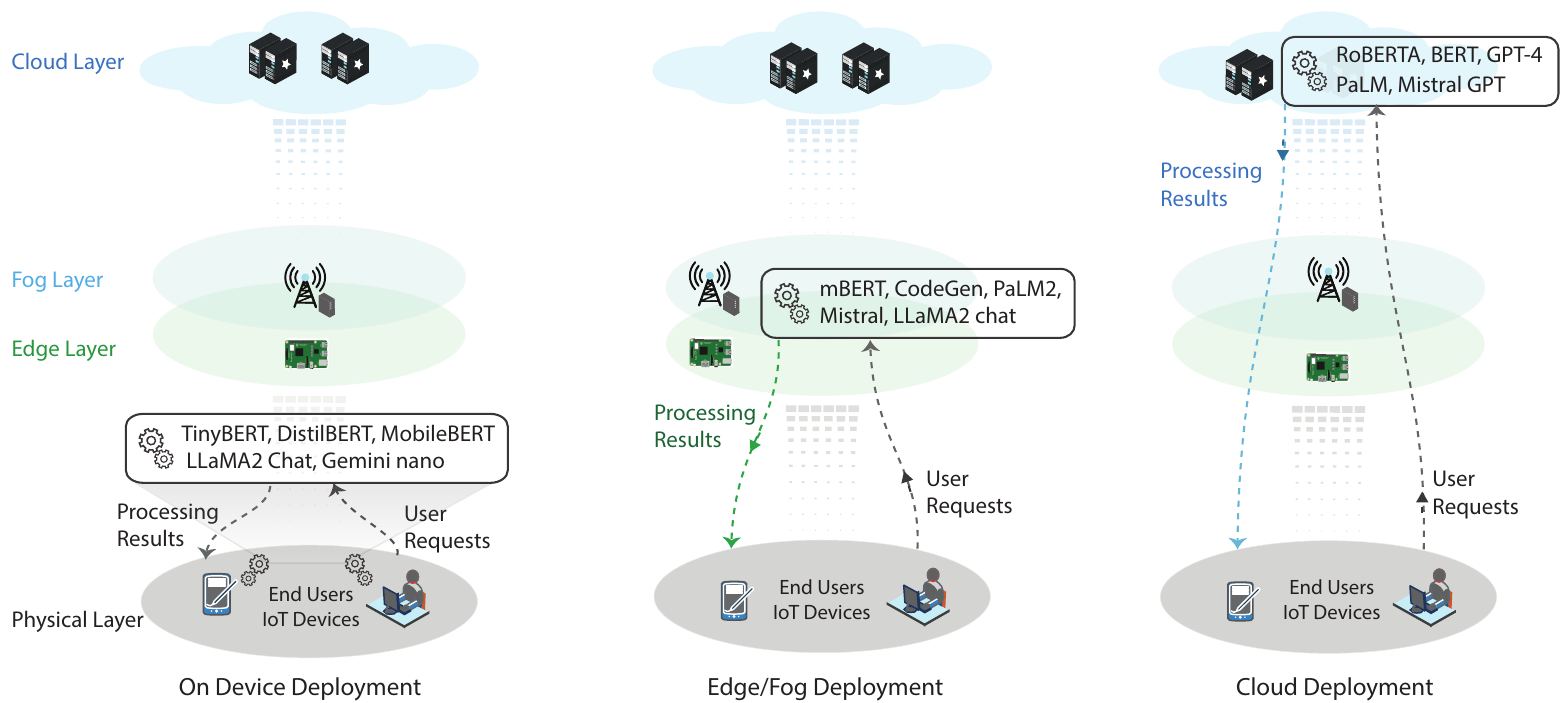}
    \caption{LLM Deployment Strategies in IoT}
    \label{llm_IoT_deploy}
\end{figure*}
\subsection{Reasoning Frameworks in LLMs}
\subsubsection{Prompting}
In LLMs, there are various prompting techniques for different tasks. These techniques include different methods used to ensure the model generates accurate and effective responses. Standard Prompting is the basic approach where an LLM receives an input and generates a response based on its pre-trained knowledge, without any added structuring. Zero-shot Prompting is similar to standard prompting but includes a structured, clear instruction to enhance context and response accuracy. Few-Shot Prompting provides a prompt with a few examples, guiding the model learn task patterns from input-output pairs, which improves performance on tasks like translation and text generation \cite{logan2021cutting}. Chain of Thought (CoT) prompting enhances LLM reasoning by encouraging step-by-step thinking, where each "thought" builds on the last. This approach helps the model break down complex tasks, improving accuracy and clarity of responses in multi-step decision-making and logical inference tasks \cite{wei2022chain}. Tree of Thought (ToT) is an advanced, branched version of Chain of Thought that explores multiple potential paths, or branches, of reasoning at each decision point. This non-linear, multi-path approach allows for a broader exploration of possible solutions, making ToT particularly useful for complex decision-making and planning tasks  \cite{yao2024tree}. Program of Thoughts (PoT) extends the CoT algorithm in a programming alike way with structured, algorithmic sequence of steps. PoT is powerful at tasks with conditional or multi-step process requiring complex problem-solving or planning abilities \cite{chen2022program}.

\subsection{Retrieval-Augmented Generation (RAG)}
    RAG is a hybrid retrieval and generation framework that enables a model to access domain-specific knowledge through two steps: first, retrieving relevant information from an external source, and then using it alongside the prompt to generate a more accurate, contextually grounded response \cite{lewis2020retrieval}. This makes RAG ideal for tasks needing specialized or current information.

\subsection{Model and Parameter Optimization}For infrastructures like IoT, which consist of resource-constrained and heterogeneous devices, LLMs are generally large and computationally expensive. Therefore, optimization techniques such as quantization (reducing the bit width of data), pruning (removing unnecessary weights), and knowledge distillation (transferring knowledge from a large model to a smaller one)  are commonly employed to reduce model size and computational costs \cite{menghani2023efficient}. In this context, techniques such as low-rank factorization and weight sharing can be employed in specific scenarios to further enhance model efficiency. Moreover, Parameter-Efficient Fine-Tuning (PEFT) methods, such as LoRA (Low-Rank Adaptation), enable more efficient training by selectively focusing on critical parameters, thereby significantly reducing computational requirements \cite{han2024parameter}.

\section{IoT-LLM Integration}

\subsection{Why IoT needs LLMs?}
IoT is an innovative technology that generates large amounts of data from heterogeneous devices and continuously processes this data to provide benefits to users, businesses, and industries. However, much of this data is raw, unstructured, and complex, making data processing challenging. In this context, LLMs have the potential to play a significant role in making IoT data more meaningful, particularly with their capabilities in natural language processing and text understanding. Most IoT data typically consists of technical and numerical information, which needs to be interpreted and presented in a user-friendly language. LLMs have the ability to analyze such data and transform it into meaningful reports, suggestions, or alerts. With IoT-LLM integration, IoT data can become more understandable and accessible, allowing managers to make more informed decisions. Furthermore, when integrated into IoT applications, LLMs can also respond to natural language queries, facilitating user interaction with the system.

On the other hand, in some specific IoT domains, the relationships between different data sources need to be determined and deeply analyzed within the context of domain knowledge. Therefore, the integration of domain knowledge and complex reasoning processes into IoT systems is essential \cite{6512846}. In this regard, LLMs can provide complex reasoning at an expert level by deeply analyzing data from various sources within a specific domain. 
For example, LLMs can evaluate data from sensors built into robotic instruments, haptic feedback devices, and real-time patient monitoring systems in a Tactile IoT system for remote robotic surgery \cite{senturk2022internet}. The LLM can help the robotic system with delicate tasks, including detecting and reacting to tissue resistance during surgery, by integrating medical domain knowledge. This can ensure accuracy and enhances operation safety.

\subsection{Why LLMs need IoT?}
LLMs need real-time, multimodal, and rich data sources to best understand more advanced meaning and context-based decisions. There is great potential to meet this need due to the heterogeneous nature of IoT. IoT in particular can provide LLMs with a continuously updated data source, allowing them to better understand dynamically changing environments and produce more accurate and up-to-date responses. For example, building temperature data from an IoT device can enable an LLM model to analyze this data and make recommendations about the energy efficiency of a building. In this way, models can become more specialized and context-sensitive.

\section{LLM-Enabled IoT 
Architectures}
Deployment of LLM models in IoT systems is a critical issue due to large parameter spaces, long inference times, and high energy consumption. Therefore, it is necessary to select the appropriate implementation environment to balance the resource demands and ensure system efficiency, specific to the task or problem. In this context, as shown in Figure \ref{llm_IoT_deploy}, LLM models can be designed with three computational architectures: on-device, edge/system, or cloud.
\subsection{On-device Deployment}
This deployment model aims to run the LLM model on devices at the physical layer. In this model, all data processing occurs locally on the device, allowing users to get results quickly and securely. The main advantage of deploying LLM on the device is that it offers the ability to protect data privacy since all data is processed locally. However, due to the limited resources of the device, large models may be problematic to run. Therefore, optimization techniques for model size and parameters may need to be applied \cite{patel2023polca,yang2024perllm,cai2024edge,yu2024edge,zhao2024edge,liu2024resource}.

\subsection{Edge/Fog Deployment}
Edge and fog computing represent deployment models that enable LLMs to be processed closer to the point of data generation. This approach facilitates local data processing within the edge and fog network. This reduces network traffic and shortens processing times by only transmitting necessary data to central servers. Edge/fog deployment reduces the dependency on centralized cloud servers, providing a faster data processing experience, especially for applications that require low latency \cite{zagar2024dynamic, zhang2024edgeshard, wu2024fedbiot, gao2024dlora}. This model is particularly well-suited for application areas such as smart cities, industrial IoT, and smart factories, where real-time data processing and low latency are critical .

\subsection{Cloud Deployment}
In LLM-IoT integration, cloud-based deployment holds significant potential, particularly for computationally intensive and multimodal IoT tasks, due to its high processing power and large-scale data processing capabilities. In such tasks, the accurate analysis of real-time data is essential for making rapid and effective operational decisions. In this regard, LLMs can offer valuable insights by interpreting complex data from IoT devices through advanced language processing and contextual understanding capabilities across large datasets. However this deployment can cause challenges in terms of latency and security due to the remote data processing \cite{mo2024iot,an2024iot,zhong2024casit,cui2024llmind}.  

\section{LLM Applications in IoT}

In this section, the existing literature is reviewed and categorized according to the specific problem areas and topics of focus. The studies primarily examine the various aspects of the integration between IoT and LLMs, highlighting the potential benefits and opportunities this integration offers.\par
\subsection{Model Deployment and Distribution}
In this section, we summarize recent studies on the use of LLM models in IoT applications across on-device, edge/fog, and cloud environments.\par

Fan et al. \cite{fan2023fate} offer the FATE-LLM , an open-source a framework for industrial applications designed to solve the efficiency challenges of federated IoT settings with different setups. In these setups, Knowledge Distillation, Offsite-tuning, several LLM architectures and PEFT algorithms are available to use. In the test experiment, ChatGLM-6B model is fine-tuned on multiple clients by LoRA or P-Tuning-v2 algorithms on AdvertiseGen  dataset. FATE-LLM achieved such performance compared to the base centralized models with only a fraction of computation cost, making it an attractive for industrial applications. 
Zagar et al. \cite{zagar2024dynamic} proposed a SpeziLLM framework that can run on edge/fog and cloud infrastructure to meet patient care and clinical research needs in IoT medical applications. In this framework, Llama2 and Gemma Phi-2 are used as base models. The authors focus on resource availability, dynamism, and data security, which are challenging in terms of resource sensitivity and data confidentiality. The proposed SpeziLLM improves response latency and cost efficiency by dynamically assigning tasks and leverages fog and cloud layers for resource-intensive processes. SpeziLLM was tested on six different applications on different mobile platforms, each with a different digital health focus, and demonstrated strong adaptability and performance. 
Yu et al. \cite{yu2024edge} emphasized that the continuous and privacy-preserving adaptation and inference requirements on edge devices are challenging to meet due to the high computational and memory demands of existing tuning techniques. To address this, they proposed Edge-LLM, incorporating three innovations: Layer-wise Unified Compression (LUC), an adaptive optimization technique based on layer sensitivity; an Efficient Layer Tuning algorithm, focusing on the most impactful layers; and a hardware scheduling algorithm to enhance efficiency. Experiments with the LLaMA backbone on MMLU and WikiText datasets show that Edge-LLM provides approximately 3x speedup and 4x memory savings compared to traditional methods, while maintaining comparable task accuracy.

\subsection{Network Management}
Zhang et al. \cite{zhang2024edgeshard} propose EdgeShard, a geographically distributed LLM architecture that optimizes model inference in IoT networks using both edge devices and cloud servers. EdgeShard brings computation closer to data sources by partitioning a single LLM model into chunks on distributed edge devices. In the proposed model, EdgeShard's goal is to improve speed and resource efficiency for long inputs and complex tasks. This model is tested on the WikiText-2 dataset using a Llama2-based framework. Experiments show that EdgeShard performs better than the cloud-edge collaborative inference method when cloud bandwidth is limited, and performs similarly when cloud bandwidth is sufficient.
Ghassemi et al. \cite{ghassemi2024multi} propose a framework combining Multi-Modal Transformer (MMT) and Reinforcement Learning (RL) to optimize beamforming in 6G wireless communication. The model first processes multi-modal inputs (images, LiDAR, radar, GPS) with MMT to predict optimal beam groups, reducing the decision space. RL then selects the best beam angle within these groups using Q-learning to adaptively optimize performance. The authors tested the approach on the DeepSense 6G dataset, which includes dynamic urban data, and found that it outperforms baseline models (MMT-only and RL-only) by improving beam prediction accuracy and system throughput.

\subsection{Task Scheduling and Planning}
Cui et al. \cite{cui2024llmind} present the LLMind framework, an intelligent IoT system with a hierarchical structure where a central Coordinator manages communication between edge LLMs, AI modules, and IoT devices. At its core, a Finite State Machine (FSM)-driven architecture controls system states like "Monitoring," "Alert," and "Response." LLMs generate Python scripts to trigger AI modules and manage IoT devices via network connections. State transition prompts guide the LLM’s decision-making based on past experiences. Experiments show that the combination of FSM control and adaptive LLM verification enables LLMind to efficiently handle complex IoT tasks with high responsiveness and accuracy.
Zhong et al. \cite{zhong2024casit} present CASIT, a collective intelligent agent system for IoT, based on multiple LLM agents that work together to solve complex tasks through cooperation. The framework includes a Memory and Summary Mechanism, allowing LLMs to efficiently process data by comparing historical data with local knowledge and chat history. In the CASIT model, ChatGLM and LLAMA models are used as the core LLMs. The authors tested CASIT using 200 sets of temperature and humidity data from five locations, and found it outperformed single LLM systems in identifying abnormal information. The system offers a new approach to information processing in IoT and provides insights for edge computing and semantic communication.

\subsection{Task Reasoning and Modelling}

Xu et al. \cite{xu2024penetrative} introduce "Penetrative AI," a framework that extends LLMs' capabilities to interact with and reason about the physical world through IoT sensors and actuators. This exploration focuses on two levels of LLMs' ability to process sensory signals and make decisions based on them. The authors demonstrate that LLMs, particularly ChatGPT, exhibit significant proficiency in using embedded world knowledge to interpret IoT sensor data and apply reasoning to tasks in the physical realm. In experiments with diverse smartphone sensor data (accelerometer, satellite, WiFi, ECG), LLMs like ChatGPT-4 and PaLM 2 successfully classified motion modes, identified environmental contexts, and monitored heart rates. These findings highlight the potential of LLMs to extend beyond traditional text-based tasks, offering new applications in cyber-physical systems and enhancing human knowledge integration in IoT.

An et al. \cite{an2024iot} propose a unifying framework called IoT-LLM to enable LLMs to understand real-world IoT tasks more effectively by using multi-modal IoT sensor data and relevant knowledge. The authors focus on enhancing the perceptive abilities of LLMs by addressing their limitations in understanding the physical world. Within the IoT-LLM framework, three steps are implemented to better adapt LLMs to IoT data: preprocessing data to make it suitable for LLMs, activating commonsense knowledge through chain-of-thought prompting, and retrieving IoT-specific information. The proposed framework was evaluated with a new benchmark set containing five different (Human Activity Recognition (HAR), Industrial, Electrocardiogram (ECG), Human Sensing, and Indoor Localization) IoT task. For this evaluation, a variety of LLM models, including GPT-3.5, GPT-4, claude-3-5-sonnet, Gemini-pro, Mistral, and LLama2, were used. Experimental results show that IoT-LLM improves the performance of existing LLMs on IoT tasks by 65\%. This study highlights the potential of LLMs to better understand IoT data and physical laws, while also suggesting new directions for future research.
Mo et al. \cite{mo2024iot} focus on the application of LLMs in interpreting multimodal data within an expanding IoT ecosystem. In this context, the authors developed a specialized language model, called IoT-LM, for processing data from various sensor modalities. This model demonstrates capabilities in areas such as human health, physical object monitoring, and smart city applications within the IoT ecosystem. A comprehensive IoT dataset called MULTIIOT was created for the developed model; this dataset contains 1.15 million instances of 12 modalities and 8 different IoT tasks. IoT-LM uses a customized multitasking adapter layer to process this multi-modal data, enabling information sharing between different tasks and modalities. The results show that IoT-LM provides a significant performance improvement in IoT classification tasks and demonstrates interactive question-answer, reasoning and dialog capabilities with IoT sensors.

Xiao et al. \cite{xiao2024efficient} propose the GIoT framework, designed for efficient and privacy-preserving analysis of semi-structured Table-QA data in IoT applications. The framework uses tables as a static memory resource for the LLM, guided by prompt engineering. It introduces Tab-PoT (Three-Stage Program of Thought), a structured prompting technique that enhances the efficiency and accuracy of Table-QA tasks. By decomposing the process into planning, conducting, and correction stages, the framework improves the model's ability to extract real-time insights from structured data queries. Additionally, optimization techniques such as token-level quantization and structured prompt management were used in the study to reduce computational costs and improve the efficiency of IoT data processing, with the aim of preserving user privacy. The GIoT framework was evaluated with LLM models like Mixtral-8x7B, Mistral-7B, and DeepSeek-67B, and tested on datasets such as WikiTableQA and TabFact. Results demonstrated the framework's scalability and robustness in privacy-sensitive, real-time IoT data analysis.

\begin{table*}[!h]
\caption{Overview of Existing Studies on IoT-LLM Integration}
\label{tblModels}
\centering
\resizebox{\textwidth}{!}{%
\begin{tabular}{|l|l|l|l|l|l|l|l|}
\hline
\textbf{\begin{tabular}[c]{@{}l@{}}Problem/Focus\end{tabular}} &
  \textbf{Reference} &
  \textbf{\begin{tabular}[c]{@{}l@{}}Model/\\ Framework\end{tabular}} &
  \textbf{Base Models} &
  \textbf{Deployment Level} &
  \textbf{Reasoning Framework} &
  \textbf{Optimization Techinques} &
  \textbf{Dataset} \\ \midrule
\multirow{4}{*}{\textbf{\begin{tabular}[c]{@{}l@{}}Task Reasoning \\ and Modelling\end{tabular}}} &
  Mo et al. \cite{mo2024iot} &
  IOT-LM &
  \begin{tabular}[c]{@{}l@{}}LLaMA-7B, LLaMA-13B, \\ GPT-4-70B\end{tabular} &
  Cloud &
  \begin{tabular}[c]{@{}l@{}}Prompt Engineering \\ (Zero-shot, Few-shot),\\ Multimodal Fusion\end{tabular} &
  N/A &
  MULTIIOT \\ \cmidrule(l){2-8} 
 &
  Hassanin et al. \cite{hassanin2025pllm}&
  IOT-LLM &
  \begin{tabular}[c]{@{}l@{}}GPT-4-175B, Claude-3.5-70B,\\ Llama2-7B, Mistral-7B, \\ Gemini-pro-67B\end{tabular} &
  Cloud &
  \begin{tabular}[c]{@{}l@{}}Prompt   Engineering\\ (Chain-of-thought), \\ RAG, \\ Multimodal Fusion\end{tabular} &
  N/A &
  \begin{tabular}[c]{@{}l@{}}MIT-BIH ECG, \\ WiFi CSI\end{tabular} \\ \cmidrule(l){2-8} 
 &
  An et al. \cite{an2024iot}&
  -&
  \begin{tabular}[c]{@{}l@{}}ChatGPT-4, \\ PaLM 2\end{tabular} &
  Cloud &
  N/A &
  N/A &
  MIT-BIH Arrhythmia \\ \cmidrule(l){2-8} 
 &
  Xiao et al. \cite{xiao2024efficient}&
  -&
  \begin{tabular}[c]{@{}l@{}}Mixtral-8x7B, Mistral-7B, \\ DeepSeek-67B, DeepSeek-7B\end{tabular} &
  Edge-Cloud &
  Prompt Engineering &
  \begin{tabular}[c]{@{}l@{}}Model optimization\\ (Quantization)\end{tabular} &
  \begin{tabular}[c]{@{}l@{}}WikiTableQA, \\ TabFact\end{tabular} \\ \midrule
\multirow{2}{*}{\textbf{\begin{tabular}[c]{@{}l@{}}Task Scheduling \\ and Planning\end{tabular}}} &
  Zhong et al. \cite{zhong2024casit} &
  CASIT &
  \begin{tabular}[c]{@{}l@{}}ChatGLM, \\ LLAMA\end{tabular} &
  Edge-Cloud &
  \begin{tabular}[c]{@{}l@{}}Role-Playing, \\ Dynamic Memory\end{tabular} &
  N/A &
  N/A \\ \cmidrule(l){2-8} 
 &
  Cui et al. \cite{cui2024llmind} &
  LLMind &
  ChatGPT &
  Edge-Cloud &
  \begin{tabular}[c]{@{}l@{}}Role-Playing, FSM\\ Dynamic Memory\end{tabular} &
  N/A &
  N/A \\ \midrule
\multirow{3}{*}{\textbf{\begin{tabular}[c]{@{}l@{}}Model Adaptation \\ and Deployment\end{tabular}}} &
 Fan et al. \cite{fan2023fate} &
  FATE-LLM &
  \begin{tabular}[c]{@{}l@{}}ChatGLM-6B, LLaMA-7B, \\ LLaMA-13B, BLOOM-7B\end{tabular} &
  Edge-Fog-Cloud &
  N/A &
  \begin{tabular}[c]{@{}l@{}}PEFT, P-Tuning-v2, \\ LoRa, Offsite tuning,\\ Knowledge Distillation\end{tabular} &
  AdvertiseGen \\ \cmidrule(l){2-8} 
 &
 Yu et al. \cite{yu2024edge} &
  EDGE-LLM &
  LLaMA-7B &
  Edge &
  N/A &
  \begin{tabular}[c]{@{}l@{}}Model Optimization (Adaptive
  \\Prunning, Quantization), \\ Hardware  Scheduling\end{tabular} &
  \begin{tabular}[c]{@{}l@{}}MMLU, \\ WikiText\end{tabular} \\ \cmidrule(l){2-8} 
 &
  Zagar et al. \cite{zagar2024dynamic} &
  SpeziLLM &
  Llama2 Gemma Phi-2 &
  Edge-Fog-Cloud &
  RAG &
  N/A &
  N/A \\ \midrule
\multirow{2}{*}{\textbf{Network Management}} &
  Zhang et al. \cite{zhang2024edgeshard} &
  EdgeShard &
  \begin{tabular}[c]{@{}l@{}}LLaMA2-7B, LLaMA2-13B, \\ LLaMA2-70B\end{tabular} &
  Edge-Cloud &
  N/A &
  Priority-based scheduling &
  N/A \\ \cmidrule(l){2-8} 
 &
  Ghassemi et al. \cite{ghassemi2024multi} &
  -&
  N/A &
  Edge-Cloud & \begin{tabular}[c]{@{}l@{}}Q-learning,\\ Multimodal Fusion \end{tabular}&
  N/A
   &
  DeepSense 6G \\ \midrule
\multirow{7}{*}{\textbf{Security and Privacy}} &
  Zheng et al. \cite{zheng2024safely} &
  FL-GLM &
  ChatGLM-6B &
  Edge-Cloud &
  N/A &
  \begin{tabular}[c]{@{}l@{}}Model Optimization\\ (Quantization),\end{tabular} &
  N/A \\ \cmidrule(l){2-8} 
 &
  Webb et al. \cite{webb2024cyber} &
  -&
  \begin{tabular}[c]{@{}l@{}}text-embedding-ada-350M,   \\ E5-large-v2-2.8B, \\ instructor-large-500M, \\ all-MiniLM-L6-v2-22M\end{tabular} &
  Cloud &
  \begin{tabular}[c]{@{}l@{}}Embedding-based retrieval, \\ RAG-based   mapping\end{tabular} &
  N/A &
  N/A \\ \cmidrule(l){2-8} 
 &
  Ye et al. \cite{ye2024detecting} &
  SLFHunter &
  ChatGPT-4.0 &
  Edge-Cloud &
  \begin{tabular}[c]{@{}l@{}}Prompt Engineering \\ (Few-shot)\end{tabular} &
  N/A &
  N/A \\ \cmidrule(l){2-8} 
 &
  Adjewa et al. \cite{adjewa2024efficient} &
  SecurityBERT &
  BERT-11B &
  Edge-Cloud
   &
  \begin{tabular}[c]{@{}l@{}}Model Optimization\\ (Quantization)\end{tabular} &
  N/A &
  EdgeIIoTset \\ \cmidrule(l){2-8} 
 &
  Rehman et. al. \cite{rehman2023let} &
  N/A &
  GPT-2 &
  Edge-Cloud &
  N/A &
  N/A &
  N/A \\ \cmidrule(l){2-8} 
 &
  Kholgh et al. \cite{kholgh2023pac} &
  PAC-GPT &
  \begin{tabular}[c]{@{}l@{}}DaVinci-175B, Babbage-1.3B, \\ variations of GPT-3\end{tabular} &
  Cloud &
  \begin{tabular}[c]{@{}l@{}}Prompt Engineering \\ (Few-Shot Learning)\end{tabular} &
  Prompt-Free Fine-Tuning &
  \begin{tabular}[c]{@{}l@{}}TON\_IoT dataset \\ (ICMP, DNS packets)\end{tabular} \\ \cmidrule(l){2-8} 
 &
  Hassanin et al. \cite{hassanin2025pllm} &
  PLLM-CS &
  Custom LLM Model &
  Cloud &
  N/A &
  N/A &
  UNSW\_NB15, TON\_IoT \\ \midrule
\multirow{7}{*}{\textbf{\begin{tabular}[c]{@{}l@{}}Service scheduling\\\& Resource allocation\end{tabular}}} &
  Wu et al. \cite{wu2024fedbiot} &
  FedBIOT &
  LLaMA-2-7B & 
  Edge-Cloud
   &
  N/A &
  \begin{tabular}[c]{@{}l@{}}Model Optimization (KD), \\ PEFT, LoRa\end{tabular} &
  \begin{tabular}[c]{@{}l@{}}MedBIoT, GSM-8K,\\ Rosetta, HumanEvalX,\\ Dolly-15K datasets.\end{tabular} \\ \cmidrule(l){2-8} 
 &
  Yang et al. \cite{yang2024perllm} &
  PerLLM &
  N/A &
  Edge-Cloud &
  N/A &
  CS-UCB &
  N/A \\ \cmidrule(l){2-8} 
 &
  Zhao et al. \cite{zhao2024edge} &
  -&
  N/A &
  Edge &
  Speculative Decoding &
  \begin{tabular}[c]{@{}l@{}}Branch and \\ Bound Scheduling\end{tabular} &
  N/A \\ \cmidrule(l){2-8} 
 &
  Liu et al. \cite{liu2024resource} &
  -&
  N/A &
  Edge &
  N/A &
  \begin{tabular}[c]{@{}l@{}}Fractional programming,\\ Concave-Convex Procedure,\\PEFT, LoRa\end{tabular} &
  N/A \\ \cmidrule(l){2-8} 
 &
  Gao et al. \cite{gao2024dlora} &
  DLoRA &
  \begin{tabular}[c]{@{}l@{}}LLaMA-7B, \\ OPT-6.7B, \\ BLOOM-7B\end{tabular} &
  Edge-Cloud &
  N/A &
  \begin{tabular}[c]{@{}l@{}}Model Optimization \\ (Quantization, Kill \& Relieve,\\ PEFT , LoRA\end{tabular} &
  \begin{tabular}[c]{@{}l@{}} BoolQ, PIQA, SIQA,\\ BoolQ, WinoGrande, OBQA\\ HellaSwag\end{tabular} \\ \cmidrule(l){2-8} 
 &
  Patel et al. \cite{patel2023polca} &
  POLCA &
  \begin{tabular}[c]{@{}l@{}}RoBERTa, \\ GPT-NeoX-20B, \\ OPT-30B, \\ BLOOM-176B\end{tabular} &
  Cloud &
  N/A &
    \begin{tabular}[c]{@{}l@{}}Power Capping, \\ Dual-thresholding, \\ Phase-aware Adjustments\\ (Quantization)\end{tabular} &
  N/A \\ \cmidrule(l){2-8} 
 &
  Yu et al. \cite{yu2024edge} &
  Edge-LLM &
  ChatGLM2-6B &
  Edge &
  N/A &
  \begin{tabular}[c]{@{}l@{}}VDF Scheduling, \\ PEFT(LoRa), \\ Model Optimization\\ (Quantization)\end{tabular} &
  LCCC dataset \\ \bottomrule
\end{tabular}%
}
\end{table*}
\subsection{Service scheduling and Resource allocation}

Gao et al. \cite{gao2024dlora} propose DLoRA, a scalable and distributed framework for integrating LoRa across cloud and user devices while maintaining robust user data privacy. DLoRA aims to enhance efficiency by using the Kill and Relieve (KR) algorithm to identify "active" and "idle" parameter modules during training. The framework targets freezing modules with minimal changes ("killing" them) to conserve computational resources, while reactivating others to preserve accuracy. The architecture allows edge devices to handle essential computations locally, offloading heavier tasks to the cloud. Additionally, quantization techniques are employed to reduce the model size. The dynamic approach of DLoRA positions it as an effective solution for scalable, privacy-conscious LLM integration in IoT environments.
Cai et al. \cite{cai2024edge} introduced Edge-LLM, a distributed training system designed to optimize fine-tuning workloads across multiple devices. The system employs the Value Density First (VDF) algorithm to minimize the computational burden on any single device. It operates by dividing the utilized model into two components: the backbone and the adapter. While the backbone remains on the primary edge device, the adapter is distributed across other layer devices. The quantized feature map is then distributed to additional devices for further processing. The system was tested on the LCCC dataset, demonstrating significant improvements in training speed, computational efficiency, and GPU utilization. The results indicate that Edge-LLM can be effective for resource-constrained, edge-based model training.

Liu et al. \cite{liu2024resource} propose a collaborative training framework that integrates mobile users with edge servers to optimize resource allocation, thereby enhancing both performance and efficiency. This approach leverages parameter-efficient fine-tuning (PEFT) methods, allowing mobile users to adjust the initial layers of the LLM while edge servers handle the more demanding latter layers. The study formulates a multi-objective optimization problem to minimize total energy consumption and delay during training. Additionally, it addresses the common issue of instability in model performance by incorporating stability enhancements into the objective function. Through a novel fractional programming technique, a stationary point for the formulated problem is achieved. Experiments show that the proposed approach maintains good consistency and dependability across a range of mobile edge computing scenarios while also drastically lowering training latency and overall energy consumption. 

Patel et al. \cite{patel2023polca} propose POLCA framework, a power oversubscription framework, to manage resources in cloud systems hosting LLM. POLCA targets oversubscription, load balancing, and phase-aware adjustments by dynamically optimizing power resources based on workload priority. The proposed framework is tested using throughput data from LLM inference clusters, including models such as RoBERTa, GPT-NeoX, OPT, and BLOOM. The results show that POLCA improves overall resource efficiency and reduces energy consumption by managing peak power loads and through frequency scaling and power capping.

Yang et al. \cite{yang2024perllm} introduced PerLLM, a personalized inference scheduling framework that enhances the deployment of large language models across edge and cloud systems by efficiently distributing multi-user requests with varied computational needs. Employing a real-time adaptive approach named as the constraint satisfaction-based Upper Confidence Bound (CS-UCB) algorithm, PerLLM schedules tasks intelligently based on resource availability. The framework assesses each task’s requirements in terms of latency and computational complexity, deciding whether to utilize edge or cloud servers accordingly. PerLLM tested with LLaMA2-7B on the edge and LLaMA2-33B on the cloud. In experimental trials deploying models such as LLaMA2-7B on the edge and LLaMA2-33B on the cloud. Experimental results showed PerLLM’s effectiveness in enhancing resource efficiency, reducing latency, and lowering power consumption
Zhao et al. \cite{zhao2024edge} present a Cooperative Inference Framework that establishes a collaboration between a small-scale language model (ssLLM) embedded in the terminal on the user side and a large-scale language model (lsLLM) located on the edge. Instead of directly sending user input to the LLM, the ssLLM generates tokens serially on the terminal, reducing resource usage and conserving computational power. These tokens are then delivered to the lsLLM on the edge for parallel verification and correction, with the Branch and Bound algorithm employed to optimize task allocation and scheduling between the ssLLM and lsLLM, minimizing latency and computational costs. After corrections, the tokens are sent back to the ssLLM as feedback, enabling an iterative process for generating new batches of tokens in the next cycle. In evaluations, this cooperative approach demonstrates superior performance compared to using the ssLLM-only or lsLLM-only systems, combining the responsiveness of the ssLLM with the accuracy of the lsLLM. Through this approach, the ssLLM reduces the computational demands on the lsLLM, which is primarily used for correction and verification. 
Wu et al. \cite{wu2024fedbiot} introduced FedBiOT, a federated learning framework designed for efficient model fine-tuning and resource use. It utilizes a bi-level architecture where clients fine-tune a lightweight adapter (the last layers of the model) using the LoRa algorithm, while the server compresses and distills the remaining layers with knowledge distillation. The updated adapter and emulator weights are iteratively exchanged between the server and clients using the FedAvg algorithm. FedBiOT reduces computational demands on clients, making it ideal for resource-constrained IoT applications.

\subsection{Security and Privacy}
Rehman et al. \cite{rehman2023let} proposed a privacy-preserving framework that enables secure use of LLMs for sensitive, edge-generated time-series data in the cloud. The framework applies adaptive differential privacy, adding noise based on context-specific sensitivity (data anonymization) to balance privacy and data utility. It uses techniques like rule-based sampling, event triggers, and lightweight edge analytics to optimize data transmission, reducing latency and bandwidth usage. The framework was tested on power consumption and solar power generation datasets, and the results showed that anonymized data retained comparable accuracy in LLM-generated responses.
Zheng et al. \cite{zheng2024safely} proposed a federated learning framework called FL-GLM for sensitive data security of end-user side clients. This approach has the characteristics of keeping critical components on client devices, offloading heavy computations to the server, and ensuring computational efficiency. In addition, FL-GLM enables real-time applications with resource-constrained clients for fast training and optimized resource utilization. The proposed framework is evaluated on understanding and summarization tasks using datasets such as SuperGLUE, CNN/DailyMail and XSum. The results show that FL-GLM performs competitively with existing benchmark models and achieves a reduction in training time. These results make FL-GLM applicable to IoT applications such as healthcare and finance where data security is paramount.
Hassanin et al. \cite{hassanin2025pllm} propose PLLM-CS framework, one of the first article in IOT-LLM security, employs several key building blocks to enhance the detection of cybersecurity threats such as fuzzer, DoS and reconnaissance attacks in both satellite and IoT environments. PLLM-CS developed an  IDS by a custom transformer architecture allowing the model to capture more subtle context-sensitive patterns required for precise threat detection. the framework tested on the UNSW\_NB 15 and TON\_IoT datasets. Experimental results show that the PLLM-CS method achieves an outstanding accuracy of 100\% on the UNSW\_NB 15 dataset.

Ye et al. \cite{ye2024detecting} developed a framework, SLFHunter, that detects command injection (CI) vulnerabilities in dynamically linked library functions (DLLFs) used in Linux-based IoT firmware.  SLFHunter provides a solution to the addressed problem by integrating static taint analysis and LLM integration. The proposed framework uses LLMs to detect sink library functions (SLFs) that can send user input to instruction execution domains without sufficient sanitization. Experimental results show that SLFHunter is faster and more accurate than existing approaches.
Webb et al. \cite{webb2024cyber} proposed a cyber attack mapping framework that links CAPEC and MITRE ATT\&CK to detect attack patterns. The framework uses CAPEC descriptions, which represent attack methods, and applies prompt engineering to connect these patterns with MITRE ATT\&CK tactics. By utilizing these CAPEC and ATT\&CK descriptions as prompts and RAG, the LLM identifies potential linkages between attack patterns and objectives. The model was evaluated on a small, hand-labeled dataset of attack pattern mappings, and the authors introduce novel metrics, such as Coverage and False Mapping Ratio (FMR), to assess mapping quality. Experimental results demonstrate high coverage with minimal false mappings, showing the framework’s effectiveness as a scalable IDS solution for 5G network security.

\begin{figure*}
    \centering
   \includegraphics[scale=.50]{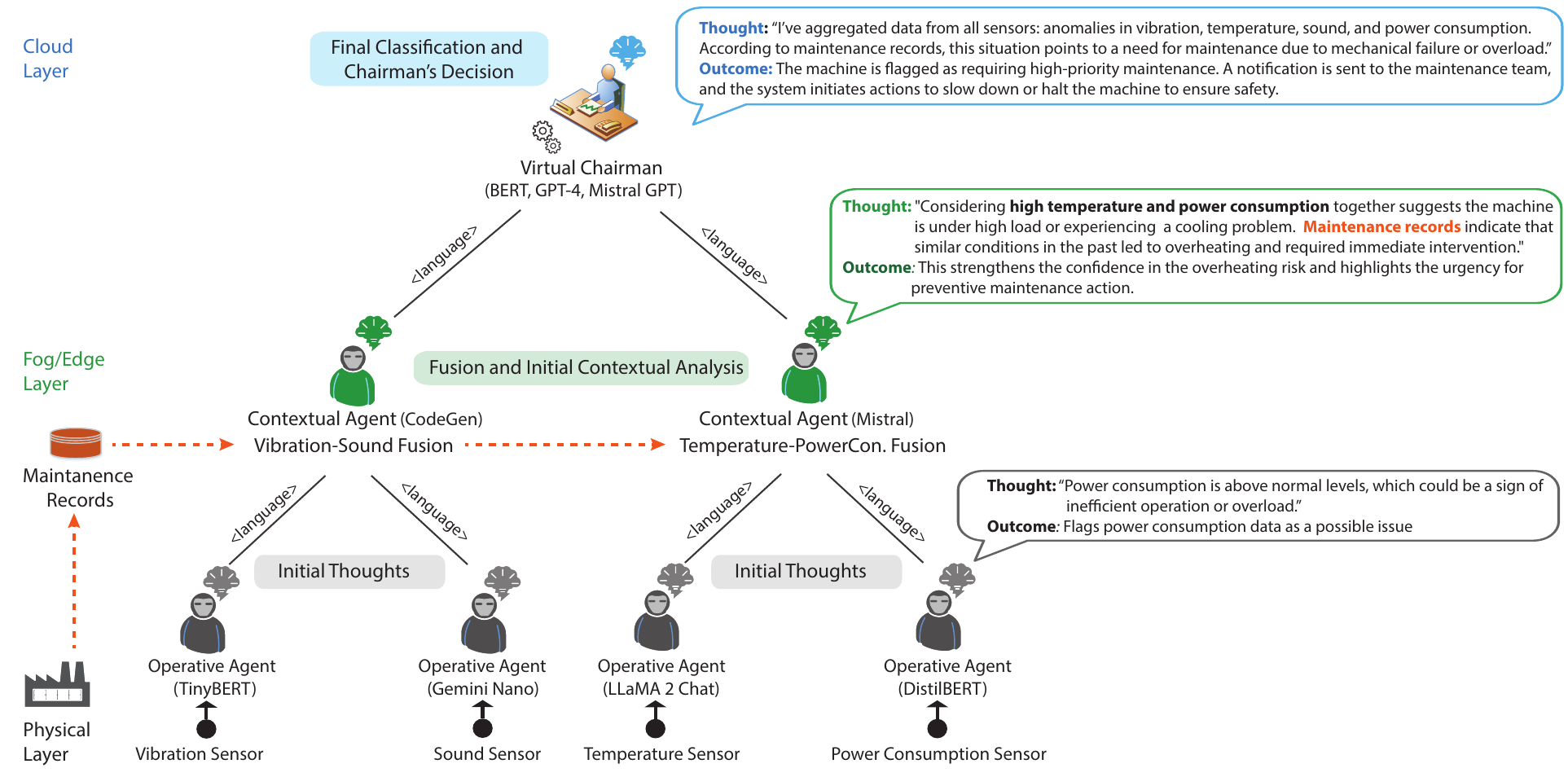}
    \caption{Proposed ToT-based Collective Intelligent System Model for IIoT}
    \label{llm_IoT_deploy}
\end{figure*}

Kholgh et al. \cite{kholgh2023pac} propose PAC-GPT framework that leverages Large Language Model (LLM) Chaining to generate synthetic network traffic supporting tasks like intrusion detection without reliance on sensitive, real-world datasets by two main components such as the Flow Generator and the Packet Generator. The Flow Generator produces the summarization of real network data. The Packet Generator, a variation of GPT-3 known as Babbage, is fine-tuned on packet data generated by DaVinci—a larger GPT-3 model—using a few-shot learning approach. In this setup, DaVinci is fed the Flow Generator’s output to produce synthetic packet examples, which are then used to fine-tune Babbage. This technique optimizes the model size, eliminating the need for costly prompt engineering while achieving similar results.

Adjewa et al. \cite{adjewa2024efficient} presents SecurityBERT, a lightweight model optimized with federated learning to detect 5G IoT network attacks. Built on the BERT backbone, SecurityBERT uses linear quantization to reduce model size, making it suitable for edge deployment. Federated learning with FedAvg aggregates model updates from multiple devices, enhancing robustness while protecting individual data privacy. Tested on the EdgeIIoTset, a 5G IoT dataset, SecurityBERT achieved high accuracy in centralized settings and comparable performance in federated scenarios. This approach provides a scalable, edge-compatible IDS solution for 5G network security, effectively addressing latency and bandwidth constraints in detecting network intrusions.

\section{Use Case: A ToT-Based Collective Intelligent System for Industrial IoT}

In this paper, we provide a system model that uses LLMs to assist in machine condition monitoring and maintenance prediction for industrial IoT applications. The Tree of Thought (ToT) reasoning framework used to enhance the complex problem-solving and cognitive capacities of LLMs is the basis of the proposed system model. At each stage in this framework, the model generates different thought paths and branches potential solutions. After these branches are evaluated against predetermined guidelines or goals, the most efficient route for the defined IoT task is selected, which allows the proposed model to derive more methodical and rational results from difficult tasks. The architectural layers of the proposed system and other components are detailed below.

\subsection{Proposed Three Tier ToT-based IIoT Architecture}
In the proposed architecture, we suggest a multi-agent collective intelligence system for IIoT that adopts the ToT approach. The following describes the information and components of each layer:

\begin{itemize} 
\item Physical Layer: This layer includes sensors that measure each parameter in an industrial working environment, capturing data such as vibration, temperature, sound, and power consumption. For our proposed model, we locate vibration, temperature, sound, and power consumption sensors. Additionally, in this layer, an operative agent is positioned to process the data from each sensor. These agents are responsible for performing basic cognitive processing to understand the status of the environmental data and prepare them for further analysis. 
\item Edge/Fog Layer: This layer contains conceptual agents that combine data from the lower layer and make inferences. These agents merge data from different sensors and perform higher-level analyses. In this process, deeper cognitive processing and inference are carried out to make more accurate predictions about machine failure. The contextual analysis performed by the agents in this layer allows for early detection of machine operational issues.  
\item Cloud Layer: In this layer, there is a virtual chairman that manages all agent relationships, coordinates the obtained results, and makes final decisions. The chairman collects all the data specific to the addressed problem, analyzes the data obtained from the fusion, and makes the final decisions regarding the machines' status. This layer also manages the prediction of maintenance needs, forecasting failures, generating early warning signals, and informing the maintenance team.
\end{itemize}
\subsubsection{Process and Data Flow}In the proposed IIoT system, data flows from the physical sensors through the edge/fog layer and finally to the cloud layer for decision-making. At the physical layer, vibration, temperature, sound, and power consumption sensors gather real-time data from machinery. Operative Agents, working independently above these sensors, receive the raw data and conduct initial processing to detect anomalies. This preliminary analysis is then forwarded to the edge/fog layer, where Conceptual Agents perform data fusion, combining insights from multiple sensors (e.g., vibration and temperature) for more accurate predictions of potential machine faults. These agents also conduct contextual analysis, enabling the system to recognize early indicators of equipment issues. Finally, the processed data reaches the cloud Layer, where the virtual Chairman consolidates all fused data and performs advanced, AI-driven analysis. The Chairman makes final decisions on machine health, predictive maintenance, and failure forecasts. If necessary, alerts with specific recommendations are sent to the maintenance team. This multi-layered approach provides an end-to-end flow for monitoring equipment health, predicting maintenance needs, and minimizing the risk of unexpected failures in industrial processes.
\section{Challenges, Open Issues and Future Directions}
In this section, the challenges that may arise in IoT-LLM integration and the research-worthy topics are presented to guide future research.

\subsection{LLM Optimization}

The biggest challenge in integrating LLMs is due to the hardware limitations of small devices making it hard to handle the large model size. While approaches such as model compression, optimization, model partitioning, and resource management can help decrease computational requirements, significant challenges persist, especially for resource-constrained devices. Current solutions, like edge-cloud collaborative devices for model partitioning or hierarchical LLM chains, often introduce latency issues that affect real-time responsiveness. Additionally, it has been observed that there are limited applications of advanced reasoning algorithms applied to the edge devices likely due to hardware limitations.

To address these computational constraints on IoT devices, future research should prioritize optimizing LLMs specifically for edge computing while enhancing their reasoning capabilities. Emerging techniques like sparse activation and dynamic model scaling could enable LLMs to operate more efficiently by selectively activating only parts of the model relevant to a given task. Hybrid optimization techniques between resource management, model compression, and task partitioning studies could result in more efficient performance in terms of computational resource usage and reduced latency. These advancements will be essential for making LLMs feasible for real-time IoT applications on low-power devices.

\subsection{Adaptation to the Dynamic Environment}
IoT environments are inherently dynamic, and models must be able to adjust to fluctuating data, device conditions, and network states. While LLMs have shown strong performance on real IoT datasets with domain-specific knowledge, adapting to rapidly changing environments remains a significant challenge. Further research is needed to explore adaptive techniques, such as dynamic memory storage and context-aware models, that could enable LLMs to better respond to changing conditions in real-time.

\subsection{IoT Specific LLM Architectures}
The integration of LLMs into IoT systems has led to the testing of various LLM architectures within IoT-specific applications. However, there is no IoT-specific LLM architecture that efficiently utilizes domain-specific data while addressing the unique constraints of IoT environments that is developed yet. Such an architecture could enhance performance by improving both the accuracy and efficiency of data processing and decision-making within IoT systems.

\subsection{Model Interpretability and Explainability}
Understanding how the LLM proceeds would increase the trust and leads to the performance enhancement applications. In that regard, Explainable AI (XAI) provides transparency into how models make decisions, which is essential for trust, accountability, and validation in LLM-IoT applications. Reviewing the presented articles, it is clear that there is yet no integration of XAI into the LLM-IOT.  Thus, XAI would be quite assisting especially in critical IoT applications, such as healthcare, autonomous vehicles, and smart city infrastructure, a lack of transparency which could lead to safety risks, operational inefficiencies, and user mistrust \cite{kok2023explainable}.

\subsection{Privacy and Security}
Privacy preservation in IoT-LLM frameworks faces significant challenges, including reliance on cloud-based computation in approaches like adaptive differential privacy and federated learning, which introduces latency and data transmission vulnerabilities. Future research should focus on lightweight, fully embedded LLMs for edge devices to eliminate cloud dependency while maintaining performance. Hybrid privacy models, integrating edge processing with selective cloud augmentation, could offer a balance between privacy, efficiency, and scalability. In addition to these privacy concerns, current security applications of LLMs in IoT are limited primarily due to constraints in available datasets for robust model training. To address these limitations, further research is needed, particularly in leveraging semi-supervised and unsupervised learning techniques, to improve the adaptability and effectiveness of LLMs in IoT security. Advancing such frameworks could yield valuable insights into identifying potential vulnerabilities and devising defensive strategies.

\section{Conclusion}
In this paper, we focus on LLM and IoT integration to understand the transformations they may create in the future. By explaining the need for these two technologies to work together, we examine their applications in the literature. In this context, we provide a comprehensive summary by reviewing recent studies in the literature based on their focus topics and problems. Based on the acquired knowledge, we introduce an innovative system model for predictive maintenance and monitoring in industrial applications, supported by LLMs. Finally, we conclude the paper by presenting the challenges that may arise in IoT-LLM integration and highlighting areas for future research.

\color{black}
\bibliographystyle{IEEEtran}
\bibliography{references}

\end{document}